# Glueball Masses in QCD$_3$

Qi-Zhou Chen,$^{a,b}$ Xiang-Qian Luo,$^{c,d,e}$ Shuo-Hong Guo$^b$, and Xi-Yan Fang$^{a,b}$

$^a$ CCAST (World Laboratory),
P.O. Box 8730, Beijing 100080, People's Republic of China

$^b$ Department of Physics, Zhongshan University, Guangzhou 510275, China

$^c$ Departamento de Física Teórica, Facultad de Ciencias,
Universidad de Zaragoza, E-50009 Zaragoza, Spain

$^d$ HLRZ, Forschungszentrum, D-52425 Jülich, Germany $^\ast$

$^e$ Deutsches Elektronen-Synchrotron DESY, D-22603 Hamburg, Germany

**Abstract**

We discuss how to extract the spectroscopy of quantum chromodynamics (QCD) in the pure gauge sector from the Hamiltonian lattice field theory approach. The recently developed truncated eigenvalue equation method is applied to the estimation of the scalar glueball $0^{++}$ and $0^{--}$ masses in the (2+1)-dimensional case. These masses reach the constant values in a scaling region as required by the renormalizability.

---

$^\ast$Mailing Address

QCD predicts the existence of glueballs, the gluonic bound states formed through strong interactions of gluons. Precise calculations of the glueball spectrum will also test decisively the validity of QCD and give guide to experimentalists in future particle searches. One of the most popular and practical techniques for the nonperturbtive determination of the glueball spectroscopy is to do numerical simulations on the Euclidean lattice. A lot of progress in this direction has been made, giving increasingly accurate estimation of the glueball masses.

Here we would like to discuss an alternative approach: the Hamiltonian formulation, which by solving the lattice Schrödinger equation can directly provide not only the glueball masses from the eigenvalues, but also the profiles, i.e., the wave functions for the ground state and excited states.

QCD in 3 dimensions (QCD$_3$) [?, ?] is a nontrivial SU(3) gauge theory, which shares a lot of important features of QCD$_4$, but is much more simplified due to lower dimensionality and superrenormalizability. Furthermore, it is hopeful that some techniques and findings in QCD$_3$ may be generalized to its 4 dimensional physical relative.

In a previous paper [?], we have investigated the long wavelength structure of the ground state of QCD$_3$, and evaluated its vacuum wave function. It has been demonstrated that our lattice vacuum wave function has the correct continuum limit and nice scaling behavior for the coefficients in the continuum vacuum wave function. This is the first step towards the understanding of the structure of the glueball and hadrons. It is worth mentioning that the ground state properties of the lattice SU(2) pure gauge models, being closely related to the confinement phenomenon, have been extensively studied both numerically [?, ?, ?] and analytically [?, ?, ?, ?, ?, ?]. What is new in [?] is the exploratory study of the realistic gauge group SU(3) using the recently developed truncated eigenvalue equation method [?, ?] and some new prescription scheme for the classification of the graphs in the wave function, which lead to rapid convergence to the continuum limit.

The purpose of this paper is to give a first estimation of the scalar glueball masses in QCD$_3$ as well as their wave functions.

Concerning the notations, truncated eigenvalue equation method and classification scheme for the graphs, we refer to [?] for further details and here just give the most relevant ones. On a discrete space and continuous time lattice with temporal gauge, the dynamics of the pure gauge system is described by

$$H = \frac{g^2}{2a} \sum_l E_l^\alpha E_l^\alpha - \frac{1}{ag^2} \sum_p Tr(U_p + U_p^\dagger - 2), \tag{1}$$

which becomes the Yang-Mills Hamiltonian with a renormalized charge $e$ in the continuum limit ($a \to 0$, or equivalently $\beta = 6/g^2 \to \infty$). The superrenormalzabilty of the theory in 2+1 dimensions implies that the bare coupling $g$ and the lattice spacing $a$ have a simple relation $g^2 = e^2 a$.

The ground state can be written in the form of $|\Omega\rangle = exp[R(U)]|0\rangle$, with $|0\rangle$ being the fluxless bare vacuum and $R(U)$ being a linear combination of gauge invariant gluonic operators such as the Wilson loops. In [?], we have illustrated how to obtain $R(U)$ and the vacuum energy $\epsilon_\Omega$ by a truncation method developed in [?], where $R(U)$ is expanded in order of graphs $R(U) = \sum_i R_i(U)$, and $R_i(U)$ is classified according to the scheme described in [?], with $i$ being the order of the graphs. Up to the third order, they are

$$R_1(U) = C_1 G_1 + h.c.,$$

$$R_2(U) = \sum_{i=1}^{6} C_{2,i} G_{2,i} + h.c.,$$

$$R_3(U) = \sum_{i=1}^{29} C_{3,i} G_{3,i} + h.c., \tag{2}$$

which graphs $G_1$, $G_{2,i}$ and $G_{3,i}$ are given in Fig. 1 (see also [?]), and the coefficients $C_1$, $C_{2,i}$ and $C_{3,i}$ are determined by solving the eigenvalue equation at the corresponding order $n$:

$$\sum_l \{[E_l, [E_l, \sum_i^n R_i(U)]] + \sum_{i+j \leq n} [E_l, R_i(U)][E_l, R_j(U)]\} - \frac{2}{g^4} \sum_p Tr(U_p + U_p^\dagger)$$

$$= \frac{2a}{g^2} \epsilon_\Omega. \tag{3}$$

To calculate the glueball masses, an important procedure is to construct appropriate states for the particles. The wave functions of these states are created by gauge invariant operators with the same quantum numbers as these glueballs acting on the vacuum state. Here for simplicity we concentrate on two scalar glueballs with quantum numbers $J^{PC} = 0^{++}$ and $J^{PC} = 0^{--}$. The wave function of a glueball is of the form

$$|F\rangle = [F(U) - \frac{\langle\Omega|F(U)|\Omega\rangle}{\langle\Omega|\Omega\rangle}]|\Omega\rangle, \qquad (4)$$

which is so chosen that it is orthogonal to the vacuum state $|\Omega\rangle$. The creation operator of the state $F(U)$ is a linear combination of a complete set of graphs with the given quantum number of the glueball. The gap or the mass of the glueball $M_{J^{PC}} = \Delta\epsilon = \epsilon_F - \epsilon_\Omega$ can be obtained by solving the Schrödinger equation

$$H|F\rangle = \epsilon_F|F\rangle, \qquad (5)$$

which is reduced to the eigenvalue equation

$$\sum_l \{[E_l, [E_l, F(U)]] + 2[E_l, F(U)][E_l, R(U)]\} = \frac{2a\Delta\epsilon}{g^2}F(U). \qquad (6)$$

Again, $F(U)$ is expanded in order of graphs as

$$F(U) = \sum_i F_i(U), \qquad (7)$$

and (??) is truncated to the $n$th order as

$$\sum_l \{[E_l, [E_l, \sum_i^n F_i(U)]] + 2\sum_{i+j\leq n}[E_l, F_i(U)][E_l, R_j(U)]\} = \frac{2a\Delta\epsilon}{g^2}\sum_i^n F(U). \qquad (8)$$

We choose the operators for the glueball $0^{++}$ as

$$F_1^{0^{++}}(U) = f_1^{0^{++}} G_1 + h.c.,$$

$$F_2^{0^{++}}(U) = \sum_{i=1}^{6} f_{2,i}^{0^{++}} G_{2,i} + h.c.,$$

$$F_3^{0^{++}}(U) = \sum_{i=1}^{29} f_{3,i}^{0^{++}} G_{3,i} + h.c., \qquad (9)$$

which are the same set of graphs $G_1$, $G_{2,i}$ and $G_{3,i}$ as those in (??) but with different coefficients. Substituting (??) into (??), we obtain 36 nonlinear equations for $f_1^{0^{++}}$, $f_{2,i}^{0^{++}}$, and $f_{3,i}^{0^{++}}$, which can be solved numerically.

The operators for the glueball $0^{--}$ are

$$F_1^{0^{--}}(U) = f_1^{0^{--}} G_1 - h.c.,$$

$$F_2^{0^{--}}(U) = f_{2,1}^{0^{--}} G_{2,1} + f_{2,3}^{0^{--}} G_{2,3} + f_{2,4}^{0^{--}} G_{2,4} - h.c.,$$

$$F_3^{0^{--}}(U) = \sum_i f_{3,i}^{0^{--}} G_{3,i}^{0^{--}} - h.c., \tag{10}$$

where the coefficients are similarly determined and $G_{3,i}^{0^{--}}$ stand for the operators with $J^{PC} = 0^{--}$ chosen from $G_{3,i}$.

For large $\beta = 6/g^2$, there should be a so called scaling region where the physical quantities become constants. Equivalently, by dimensional analysis, in the lattice unit the inverse correlation length or the dimensionless masses $aM_{J^{PC}}$ should be proportional to $g^2$, i.e.,

$$\frac{aM_{J^{PC}}}{g^2} \to \frac{M_{J^{PC}}}{e^2} = const., \tag{11}$$

as required by the renormalizability of the theory.

The results for $aM_{0^{++}}/g^2$ and $aM_{0^{--}}/g^2$ are shown in Fig. 2. As one sees, there exists indeed a scaling window $\beta \in [5,8)$ for $0^{++}$ and $\beta \in (6,12]$ for $0^{--}$ where we can extract their masses (indicated by the dot lines in Fig. 2). From the data on the dot lines, we get

$$\frac{M_{0^{--}}}{M_{0^{++}}} \approx 1.6989. \tag{12}$$

To summarize, using the techniques and information on the vacuum wave function in [?], and solving the lattice Schrödinger equation truncated to the third order, we have obtained the glueball wave functions and masses in $QCD_3$. The results for the mass spectrum show sensible scaling behavior in some intermediate or weak coupling regions. (In comparison, the $QCD_4$ glueball masses from the Monte Carlo simulations in the literature are extracted within $\beta \in [5.9, 6.2]$.) Of course, when approaching the continuum limit $a \to 0$ or equivalently $\beta \to \infty$, as the correlation lengths in the lattice units will be divergent, a conventional wisdom is to include higher and higher orders of graphs to better represent the glueballs. This is quite similar to the situation in the Monte Carlo simulation on the Euclidean lattice where larger and larger lattices are required in the weak coupling regime. It may also be possible to improve the scaling using a better graphic classification (mixing) scheme. This issue and the higher order effects on the physical observables will be further investigated.

Q.Z.C., S.H.G. and X.Y.F. were supported by the Doctoral Program Foundation of Institute of Higher Education, the People's Republic of China and the Advanced Research Center of Zhongshan University, Hong Kong. X.Q.L. thanks DESY for support.

**Figure Captions**

Fig. 1.1. First order graph in $R(U)$.

Fig. 1.2. Second order graphs in $R(U)$.

Fig. 1.3(a). Third order graphs in $R(U)$.

Fig. 1.3(b). Third order graphs (continuation of Fig. 1.3(a)) in $R(U)$.

Fig. 2. $aM_{0++}/g^2$ and $aM_{0--}/g^2$ as a function of $\beta$. The dot lines indicate the corresponding mean values in the scaling region.

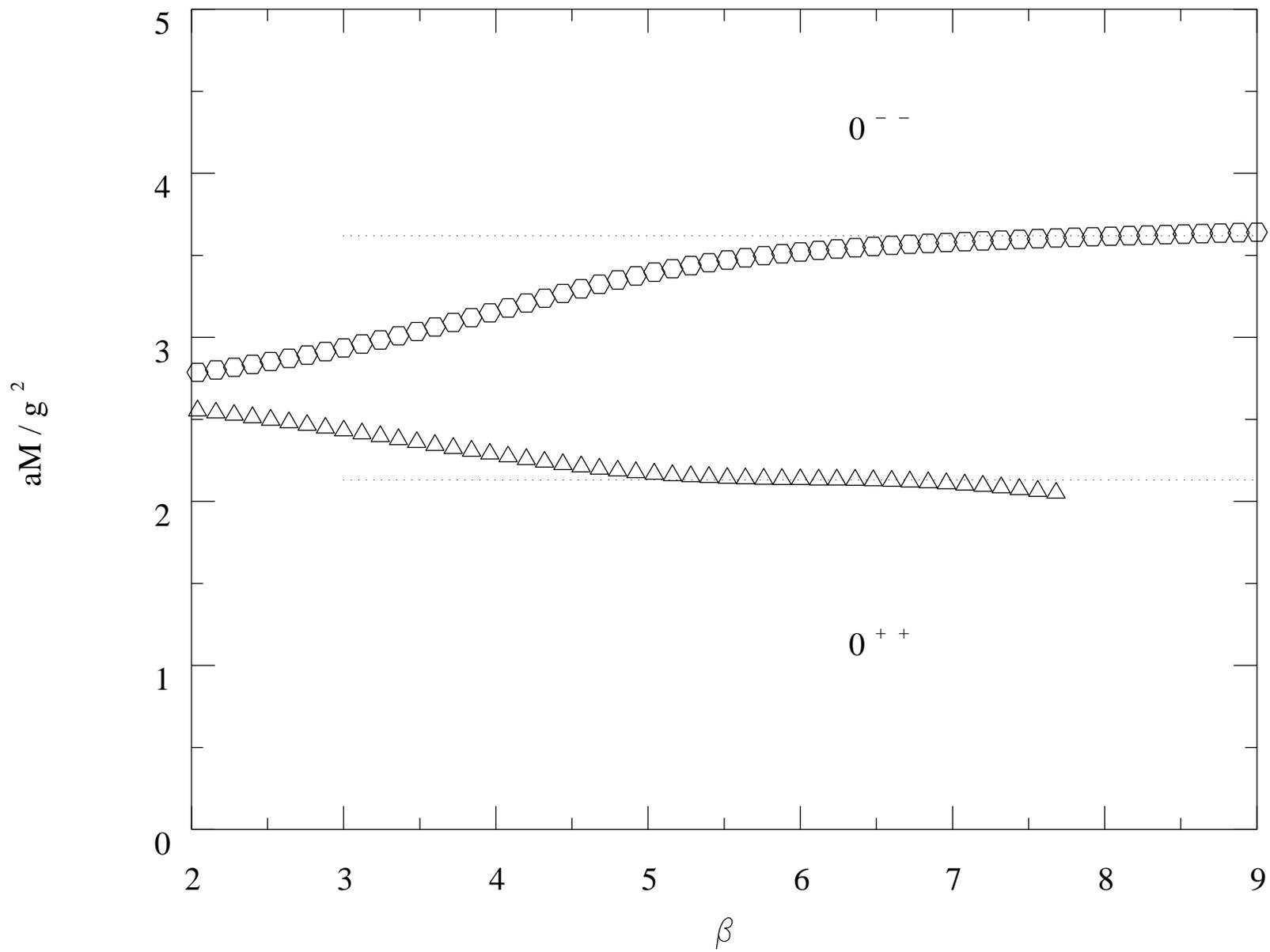

Fig. 2

$G_1$ 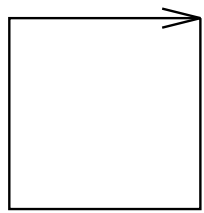

Fig. 1.1

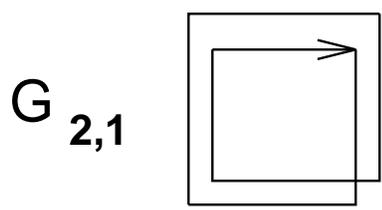 $G_{2,1}$

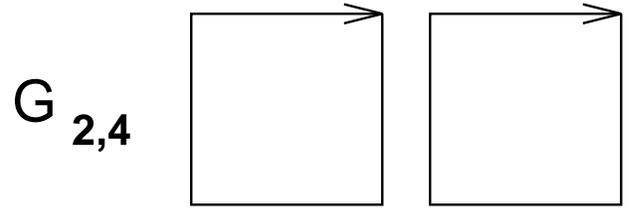 $G_{2,4}$

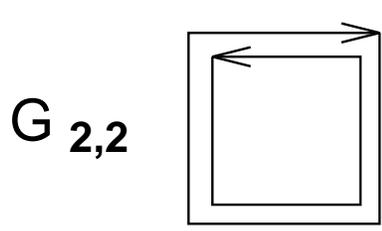 $G_{2,2}$

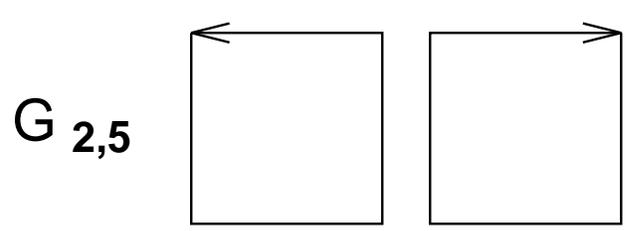 $G_{2,5}$

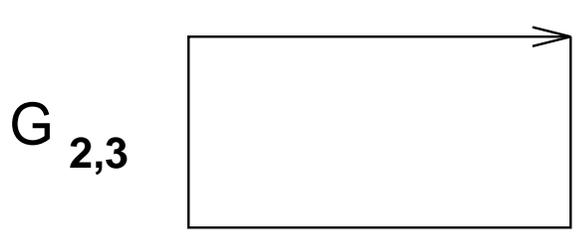 $G_{2,3}$

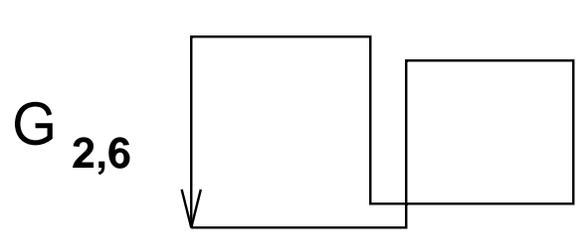 $G_{2,6}$

Fig. 1.2

G <sub>3,1</sub>

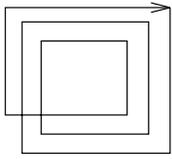

G <sub>3,2</sub>

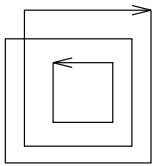

G <sub>3,3</sub>

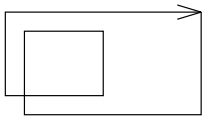

G <sub>3,4</sub>

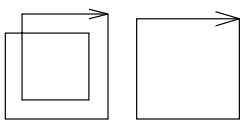

G <sub>3,5</sub>

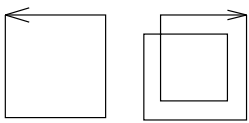

G <sub>3,6</sub>

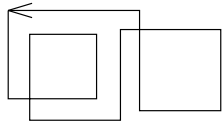

G <sub>3,7</sub>

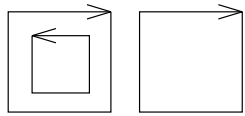

G <sub>3,8</sub>

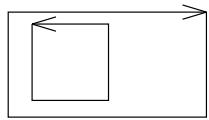

G <sub>3,9</sub>

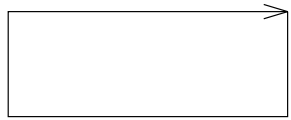

G <sub>3,10</sub>

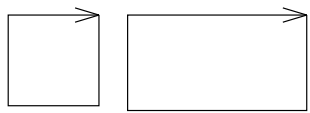

G <sub>3,11</sub>

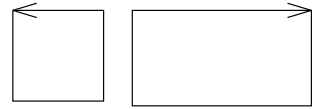

G <sub>3,12</sub>

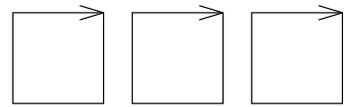

G <sub>3,13</sub>

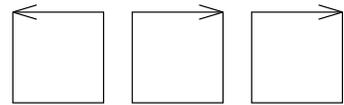

G <sub>3,14</sub>

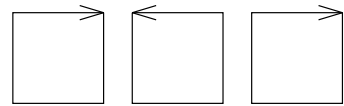

G <sub>3,15</sub>

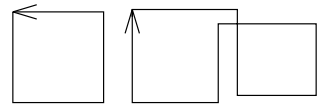

Fig. 1.3(a)

G_{3,16}

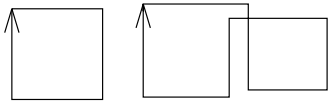

G_{3,17}

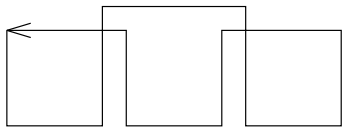

G_{3,18}

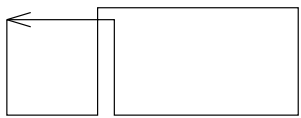

G_{3,19}

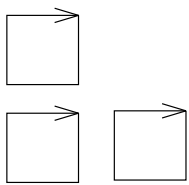

G_{3,20}

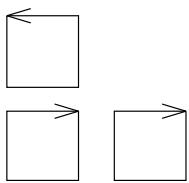

G_{3,21}

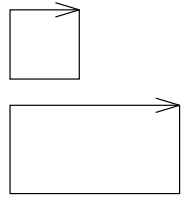

G_{3,22}

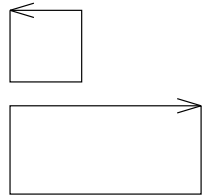

G_{3,23}

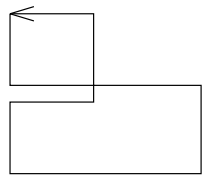

G_{3,24}

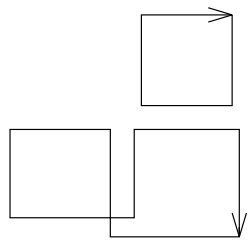

G_{3,25}

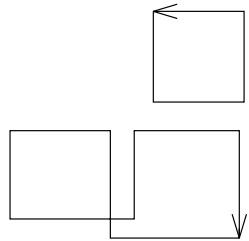

G_{3,26}

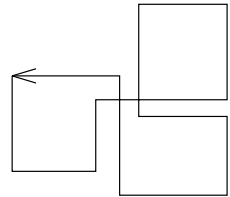

G_{3,27}

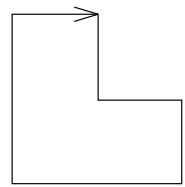

G_{3,28}

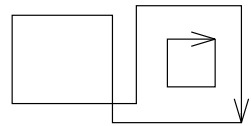

G_{3,29}

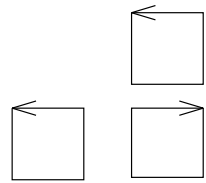

Fig. 1.3(b)